\documentclass[aps,showpacs,twocolumn]{revtex4-2}
\usepackage[utf8]{inputenc}
\usepackage{amsfonts}
\usepackage{amssymb}
\usepackage{amsmath}
\usepackage{graphicx}
\usepackage{appendix}
\usepackage{hyperref}
\usepackage{color}
\usepackage{orcidlink}

\hypersetup{hypertex=true, colorlinks=true, linkcolor=blue, urlcolor=blue, citecolor=blue}

\begin{document}

\title{Emergent one-dimensional Bloch oscillations in coupled Ising layers}
\author{E. S. Ma\, \orcidlink{0000-0002-7314-991X} }
\author{Z. Song\, \orcidlink{0000-0002-3315-4589} }
\email{songtc@nankai.edu.cn}
\affiliation{School of Physics, Nankai University, Tianjin 300071, China}
\begin{abstract}
Bloch oscillations typically arise in periodic systems subjected to a
constant force generated by a potential gradient. Here, we study domain-wall
dynamics in coupled Ising layers under asymmetric longitudinal fields. The
interlayer coupling is of Heisenberg type and acts as an effective
transverse coupling in the emergent pseudospin representation. We show that
the rung pseudospin structure gives rise to kinetic constraints, restricting
the many-body dynamics to an emergent one-dimensional quasiparticle subspace
within the original two-dimensional system. Furthermore, the imbalance
between the longitudinal fields of the two layers induces an effective
linear potential for the emergent domain-wall excitations, leading to their
confinement between two oppositely polarized ferromagnetic domains.
Remarkably, the resulting Wannier--Stark ladder of domain-wall excitations
gives rise to Bloch oscillations in the absence of a conventional tilted
potential, revealing a distinct mechanism for coherent quantum dynamics.
Numerical simulations of time-dependent observables confirm the analytical
predictions.
\end{abstract}

\maketitle




\section{Introduction}

\label{sec:introduction}

Bloch oscillations are a fundamental manifestation of quantum dynamics in
periodic systems subjected to a constant force. Originally proposed for
electrons in crystalline lattices, a constant force drives the crystal
momentum across the Brillouin zone and leads to periodic motion rather than
uniform acceleration~\cite{Bloch1929,Zener1934}. In the energy
representation, this dynamics is associated with the formation of an equally
spaced Wannier--Stark ladder~\cite{Wannier1960,Wannier1962}. Bloch
oscillations and Wannier--Stark localization have been extensively
investigated in semiconductor superlattices and optical lattices, where they
have been observed experimentally in both solid-state and ultracold-atom
systems~\cite%
{Feldmann1992,Waschke1993,BenDahan1996,Anderson1998,Morsch2001,Morsch2006}.
These systems provide a paradigmatic setting in which coherent quantum
dynamics emerges from the competition between tunneling and a spatially
linear potential.

The realization of Bloch oscillations in interacting quantum systems is
considerably more subtle. Interactions can modify the quasiparticle
dispersion, couple different excitation sectors, and eventually destroy
coherent oscillatory motion. Nevertheless, Bloch oscillations can survive in
strongly correlated systems when suitable quasiparticles remain well
defined. It has recently been established that subjecting a bound pair to a
linear potential induces periodic dynamics, even within a correlated system 
\cite%
{Khomeriki2010,Longhi2011,Longhi2012,Corrielli2013,Lin2014,ZhangXZ2016,Zhang2024, ma2026pairing}%
. In addition, a theorem for a broad class of systems, including strongly
correlated and non-Hermitian systems with conserved particle number, has
been rigorously proven in quantum many-body settings \cite%
{Zhang2025Bloch,Zhang2015formation,Zhang2024extended}. On the other hand,
domain walls in anisotropic spin chains can play the role of charged
particles in a lattice, with a magnetic field providing an effective force.
Bloch oscillations of magnetic solitons were theoretically predicted in
anisotropic spin-$1/2$ chains~\cite{Kyriakidis1998}, and magnetic Bloch
oscillations of domain walls were subsequently observed in a near-Ising
ferromagnetic chain using neutron scattering~\cite{Hansen2022}. More
recently, Bloch oscillations have also been studied in strongly interacting
spin systems, including the spin-$1/2$ XXZ chain~\cite{Liu2021}, as well as
in interacting many-body systems subject to Stark potentials~\cite%
{VanRegemortel2020}. These developments raise an interesting question: can
coherent Bloch oscillations emerge from an interacting many-body system in
which both the effective quasiparticle and the corresponding linear
potential are generated by internal degrees of freedom?

Here, we address this question in a system of two coupled Ising layers.
Unlike conventional realizations, where the force responsible for Bloch
oscillations is imposed externally along a fixed spatial direction, we show
that an effective one-dimensional Wannier--Stark problem can emerge
intrinsically from the internal structure of the coupled layers. The two
Ising layers are coupled through a Heisenberg-type interaction and are
subjected to unequal longitudinal fields. The resulting local rung Hilbert
space possesses a reducible pseudospin structure, consisting of an active
spin-$1/2$ sector and two inert singlet sectors. This structure induces
kinetic constraints and fragments the full Hilbert space into dynamically
disconnected sectors. Within each nontrivial sector, the dynamics can be
mapped onto an emergent transverse-field Ising model, providing an effective
one-dimensional description embedded in the original two-dimensional system.

A particularly important consequence of this emergent description is that
the imbalance between the longitudinal fields of the two layers acts as an
effective force on the domain-wall excitations. In the domain-wall
representation, the field imbalance generates a linear potential and
consequently an equally spaced Wannier--Stark ladder. The transverse
interlayer coupling provides hopping between neighboring Wannier--Stark
states. The resulting effective model is therefore a tight-binding problem
in a linear potential, although no conventional tilted potential is applied
to the original two-dimensional system. This mechanism leads to coherent
Bloch oscillations of the domain-wall excitations, with a Bloch period
determined by the longitudinal-field imbalance.

The mechanism revealed here is distinct from conventional Bloch oscillations
of single particles in tilted lattices and from previously studied magnetic
Bloch oscillations in single Ising chains \cite%
{zhang2024dynamics,Zhang2024magnetic, Ma2026dynamic}. In our case, the
effective lattice on which the domain-wall excitation propagates emerges
from a fragmented many-body Hilbert space, while the effective force
originates from the imbalance between two internal layers. Thus, the spatial
coordinate relevant for the Wannier--Stark problem is not the original
lattice coordinate of a single-particle system, but the position of an
emergent many-body excitation within a dynamically disconnected sector. This
provides a route toward realizing one-dimensional Wannier--Stark physics
within a higher-dimensional interacting system and illustrates how kinetic
constraints and internal degrees of freedom can generate coherent quantum
dynamics that is not apparent from the original lattice geometry.

In this work, we derive the emergent pseudospin structure and the
corresponding Hilbert-space fragmentation, construct the effective
transverse-field Ising sectors, and identify the domain-wall Wannier--Stark
ladder. We obtain analytical expressions for the corresponding
Wannier--Stark eigenstates and propagator and show that the domain-wall
dynamics is periodic with a Bloch period determined by the field imbalance.
Numerical simulations for both localized and Gaussian wave-packet initial
states confirm the analytical predictions and demonstrate clear Bloch
oscillations. Our results establish an emergent mechanism for Bloch
oscillations in coupled Ising layers and provide a new perspective on the
interplay between Hilbert-space structure, quasiparticle confinement, and
coherent many-body dynamics.

The remainder of this paper is organized as follows. In Sec.~\ref{Model and
HSF}, we introduce the coupled Ising-layer model and reveal the emergent
pseudospin structure of the local rung Hilbert space. In Sec.~\ref{Emergent
transverse-field Ising sectors}, we discuss the resulting Hilbert-space
fragmentation and construct the effective transverse-field Ising sectors. In
Sec.~\ref{Wannier-Stark ladder}, we focus on the domain-wall dynamics within
the effective one-dimensional sectors and derive the corresponding
Wannier--Stark ladder and analytical propagator. In Sec.~\ref{Bloch
oscillations}, we present numerical results for localized and Gaussian
wave-packet initial states and demonstrate the resulting Bloch oscillations.
Finally, Sec.~\ref{conclusions} summarizes our main results and discusses
their implications.

\section{Model Hamiltonian and pseudospin representation}

\label{Model and HSF}

The system considered here consists of two identical Ising layers, referred
to as layer 1 and layer 2, each subjected to its own longitudinal magnetic
field. The two layers are coupled through a Heisenberg-type interaction. The
Hamiltonian is given by%
\begin{equation}
H=\sum_{\lambda =1,2}H_{\lambda }+J\sum_{j=1}^{N}\left( \boldsymbol{\sigma }%
_{1,j}\cdot \boldsymbol{\sigma }_{2,j}+1\right) ,  \label{H}
\end{equation}%
where%
\begin{equation}
H_{\lambda }=\sum_{\left\langle i,j\right\rangle }^{N}\sigma _{\lambda
,i}^{z}\sigma _{\lambda ,j}^{z}+h_{\lambda }\sum_{j=1}^{N}\sigma _{\lambda
,j}^{z},  \label{H0}
\end{equation}%
describes the longitudinal-field Ising Hamiltonian for layer $\lambda $. The
second term represents the rung Heisenberg coupling between the two layers.
Here, $\sigma _{\lambda ,j}^{\alpha }$\ ($\lambda =1,2;\alpha =x,y,z$)
denotes the Pauli operator at site $j$ of layer $\lambda $. Throughout this
work, the Ising coupling strength is set to unity. The parameter $J$ denotes
the interlayer coupling strength, while $h_{\lambda }$ represents the
longitudinal field applied to layer $\lambda $. The constant shift does not
affect the dynamics but is introduced for convenience.\ The imbalance
between the longitudinal fields of the two layers plays a crucial role in
generating an effective potential for the emergent kink excitations.

We introduce a set of rung pseudospin operators 
\begin{eqnarray}
\tau _{j}^{+} &=&\left( \tau _{j}^{-}\right) ^{\dag }=\sigma
_{1,j}^{+}\sigma _{2.j}^{-},  \notag \\
\tau _{j}^{z} &=&\left[ \tau _{j}^{+},\tau _{j}^{-}\right] =\frac{1}{2}%
\left( \sigma _{1,j}^{z}-\sigma _{1,j}^{z}\right) ,  \notag \\
\tau _{j}^{x} &=&\tau _{j}^{+}+\tau _{j}^{-},
\end{eqnarray}%
which are pseudospin operators, obeying the SU(2) Lie algebra, i.e., $[\tau
_{j}^{+},\tau _{l}^{-}]=\delta _{jl}\tau _{j}^{z}$, and $[\tau _{j}^{z},\tau
_{l}^{\pm }]=\pm 2\delta _{jl}\tau _{j}^{\pm }$.

\begin{figure}[tbp]
\centering
\includegraphics[width=0.4\textwidth]{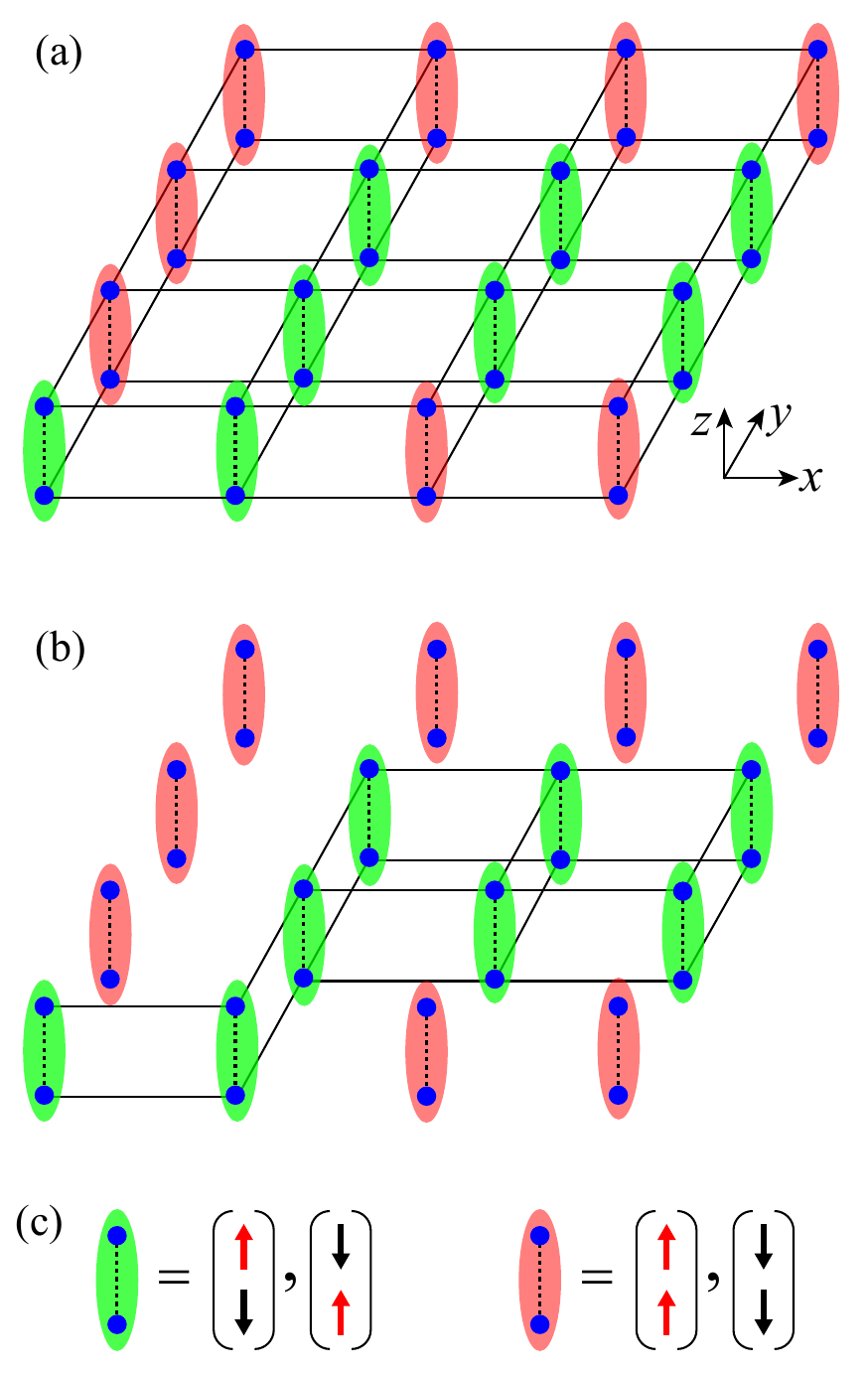}
\caption{Schematic illustration of the coupled Ising layers described by the
Hamiltonian in Eq.~(\protect\ref{H}). (a) Square-lattice model consisting of
inert (red-shadowed) and active (green-shadowed) rungs. (b) The same lattice
as in (a), where each inert rung is isolated, while the interactions within
the active rungs remain effective due to the dynamical constraints imposed
by the Hamiltonian. (c) Definitions of the active and inert rung states,
where the red and black arrows denote the eigenstates of the corresponding $%
\protect\sigma_{\protect\lambda, l}^{z}$ operators with eigenvalues $1$ and $%
-1$, respectively.}
\label{Fig1}
\end{figure}

For the $j$th rung, the local Hilbert space is spanned by 
\begin{equation}
\mathcal{H}_{j}=\mathrm{span}\left\{ \left\vert \uparrow \downarrow
\right\rangle _{j},\left\vert \downarrow \uparrow \right\rangle
_{j},\left\vert \uparrow \uparrow \right\rangle _{j},\left\vert \downarrow
\downarrow \right\rangle _{j}\right\} .
\end{equation}%
With the rung pseudospin operators $\tau _{j}^{\pm }$ the states $\left\{
\left\vert \uparrow \downarrow \right\rangle _{j},\left\vert \downarrow
\uparrow \right\rangle _{j}\right\} $ are dynamically coupled through the
pseudospin-flip processes, $\tau _{j}^{+}\left\vert \downarrow \uparrow
\right\rangle _{j}=\left\vert \uparrow \downarrow \right\rangle _{j}$ and $%
\tau _{j}^{-}\left\vert \uparrow \downarrow \right\rangle _{j}=\left\vert
\downarrow \uparrow \right\rangle _{j}$. We therefore refer to these states
as \emph{active rung states}. In contrast, $\tau _{j}^{\pm }\left\vert
\uparrow \uparrow \right\rangle _{j}=\tau _{j}^{\pm }\left\vert \downarrow
\downarrow \right\rangle _{j}=0$, so that the states $\left\{ \left\vert
\uparrow \uparrow \right\rangle _{j},\left\vert \downarrow \downarrow
\right\rangle _{j}\right\} $ are dynamically isolated with respect to the
emergent pseudospin dynamics. We refer to them as \emph{inert rung states}.
Thus, the four-dimensional rung Hilbert space decomposes into one active
spin-$1/2$ sector and two inert singlet sectors, 
\begin{equation}
\mathcal{H}_{j}=\mathcal{H}_{j}^{\mathrm{active}}\oplus \mathcal{H}_{j}^{%
\mathrm{inert},1}\oplus \mathcal{H}_{j}^{\mathrm{inert},2},
\end{equation}%
corresponding to the reducible representation $\frac{1}{2}\oplus 0\oplus 0$.
This emergent pseudospin structure enables an effective one-dimensional
description of the quasiparticle dynamics within the two-dimensional system,
which supports Wannier--Stark ladders and associated Bloch oscillations.
Fig. \ref{Fig1}(a) and (b) illustrate the system on a square lattice.

\section{Emergent transverse-field Ising sectors}

\label{Emergent transverse-field Ising sectors}

The reducible structure of the rung pseudospin Hilbert space provides a
natural mechanism for Hilbert-space fragmentation. Specifically, the
spin-(1/2) sector, spanned by ($\left\vert \uparrow \downarrow \right\rangle
_{l}$, $\left\vert \downarrow \uparrow \right\rangle _{l}$), is dynamically
coupled by the interlayer Heisenberg exchange, whereas the two trivial
sectors ($\left\vert \uparrow \uparrow \right\rangle _{l}$) and ($%
\left\vert\downarrow \downarrow \right\rangle _{l}$) remain inert under the
pseudospin operations.\ As a consequence, the full Hilbert space decomposes
into dynamically disconnected sectors. We will show that, within each
nontrivial pseudospin sector, the effective Hamiltonian can be mapped onto a
transverse-field Ising model, where the interlayer Heisenberg coupling acts
as an emergent transverse field.

We first discuss the decomposition of the Hilbert space induced by the
reducible rung pseudospin structure. A generic basis state of the full
Hilbert space can be written as 
\begin{equation}
\left\vert \psi \right\rangle =\left\vert \varphi \right\rangle _{\text{%
\textrm{A}}}\left\vert \varphi \right\rangle _{\text{\textrm{B}}},
\end{equation}%
where A and B denote the sets of active and inert rungs, containing $N_{%
\text{\textrm{A}}}$ and $N_{\text{\textrm{B}}}$ rungs respectively. For a
rung $l\in A$, the local state is restricted to the pseudospin-1/2 subspace, 
$\left\{ \left\vert {\uparrow \downarrow }\right\rangle {_{l},}\left\vert {%
\downarrow \uparrow }\right\rangle {_{l}}\right\} $, whereas for a rung $%
l\in B$, the local state belongs to the inert subspace, $\left\{ \left\vert
\uparrow \uparrow \right\rangle _{l},\left\vert \downarrow \downarrow
\right\rangle _{l}\right\} $. Therefore, different choices of the B-type
rungs define disconnected Hilbert-space sectors. For a system with $N$ rungs
and $N_{\text{\textrm{B}}}$ inert rungs, the total number of sectors is 
\begin{equation}
N_{\text{\textrm{sec}}}=\left( 
\begin{array}{c}
N \\ 
N_{\text{\textrm{B}}}%
\end{array}%
\right) 2^{N_{\text{\textrm{B}}}},
\end{equation}%
where the combinatorial factor accounts for the possible choices of inert
rungs, while the factor $2^{N_{\text{\textrm{B}}}}$ arises from the two
possible local configurations of each inert rung.

In each fragmented sector, the dynamics is governed solely by the active
pseudospin-1/2 degrees of freedom. The corresponding effective Hamiltonian
on the $N_{\text{\textrm{A}}}$ active rungs is given by 
\begin{equation}
H_{\text{\textrm{A}}}=2\sum_{\left\langle i,j\right\rangle \in A}\tau
_{i}^{z}\tau _{j}^{z}+\left( h_{1}-h_{2}\right) \sum_{j\in A}\tau _{j}^{z}+%
\frac{J}{2}\sum_{j\in A}\tau _{j}^{x},
\end{equation}%
which is an emergent transverse-field Ising model with an additional
longitudinal field induced by the imbalance between the two layers. We
emphasize that the lattice geometry of $H_{\text{\textrm{A}}}$ is not fixed
and can vary depending on the choice of the fragmented sector. In the
following, we focus on the simplest case, namely an effective
one-dimensional chain, which allows an analytical treatment. For
convenience, we rewrite the effective Hamiltonian $H_{\text{\textrm{A}}}$ in
the form of a one-dimensional transverse-field Ising model 
\begin{equation}
H_{\text{\textrm{eff}}}=2\sum_{j=1}^{N_{\text{\textrm{A}}}-1}\tau
_{j}^{z}\tau _{j+1}^{z}+\left( h_{1}-h_{2}\right) \sum_{j=1}^{N_{\text{%
\textrm{A}}}}\tau _{j}^{z}+\frac{J}{2}\sum_{j=1}^{N_{\text{\textrm{A}}}}\tau
_{j}^{x},
\end{equation}%
which provides the basis for the subsequent analysis of domain-wall dynamics.

\begin{figure}[tbh]
\centering
\includegraphics[width=0.4\textwidth]{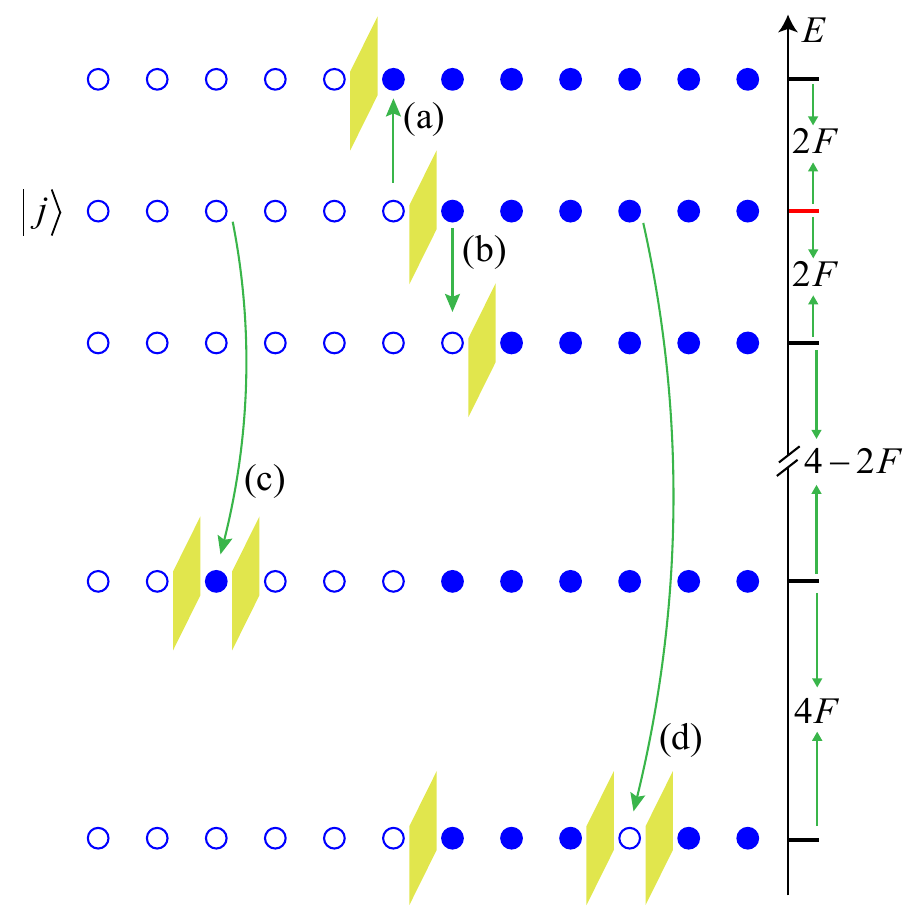}
\caption{Possible dynamical processes and the corresponding energy
transitions for a one-dimensional single-domain-wall state $\left\vert
j\right\rangle$, defined in Eq.~(\protect\ref{one-domain}), driven by $H_{%
\mathrm{T}}$ in Eq.~(\protect\ref{HT}). The open and filled blue circles
denote the active states $\left\vert \downarrow \uparrow \right\rangle _{l}$
and $\left\vert \uparrow \downarrow \right\rangle _{l}$, respectively. (a)
and (b) illustrate processes in which one active state is transformed into
the other, causing the domain wall to translate by one site, with the
corresponding energy change under $H_{\mathrm{v}}$ in Eq.~(\protect\ref{Hv})
being $\pm 2F$. (c) and (d) illustrate processes that generate two
additional domain walls, resulting in energy changes of $4\mp 2F$. Under the
condition $F,T\ll 1$, the processes in (c) and (d) are suppressed because of
their large energy costs and consequently low transition probabilities,
leaving the processes in (a) and (b) to dominate the dynamics.}
\label{Fig2}
\end{figure}

\begin{figure*}[tbh]
\centering
\includegraphics[width=1\textwidth]{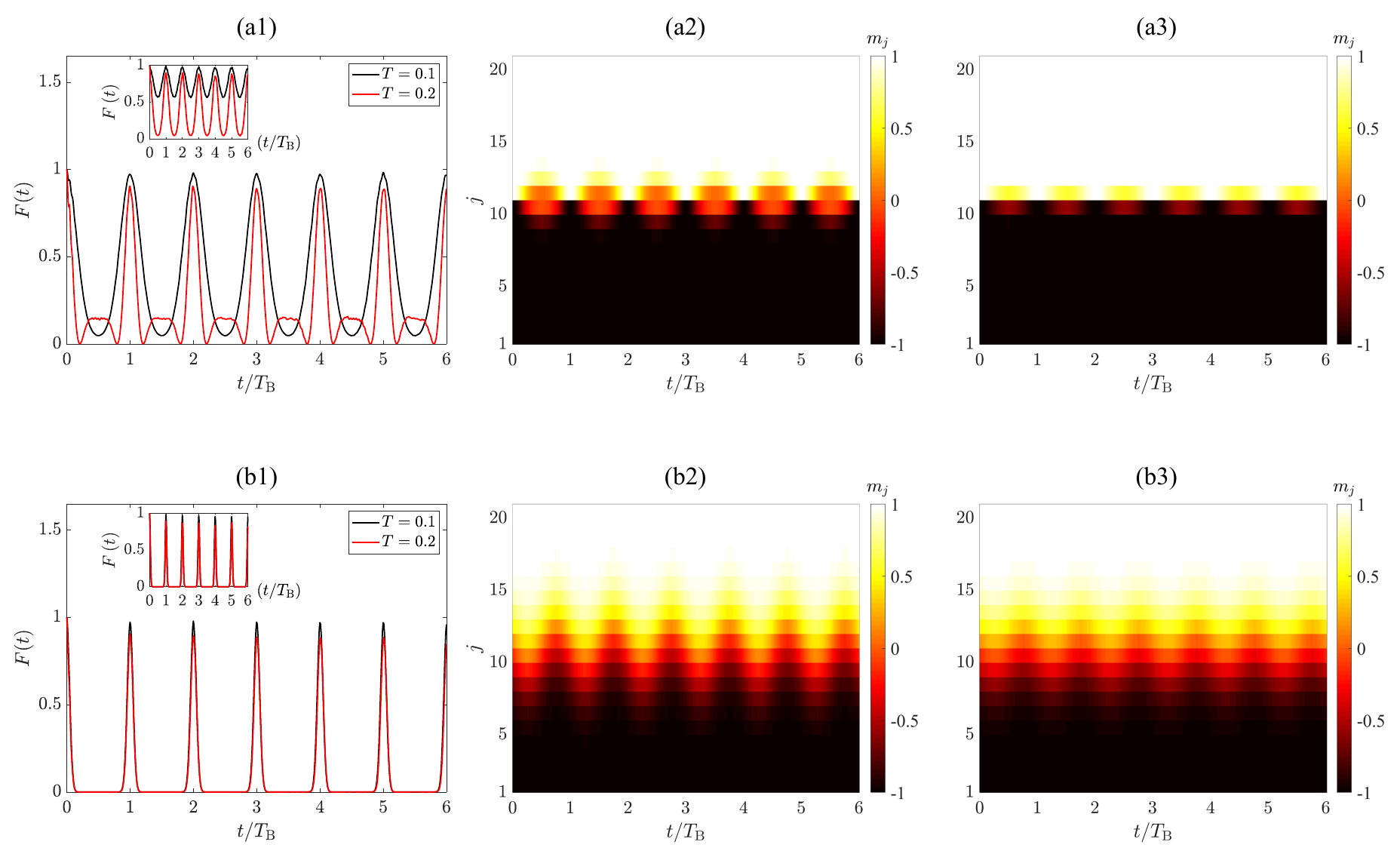}
\caption{Plots of the fidelity, defined in Eq.~(\protect\ref{F}), and the
local magnetization, defined in Eq.~(\protect\ref{m_j}), for two different
initial states in one-dimensional systems described by Eq.~(\protect\ref%
{H_eff}). Panels (a1) and (b1) show the fidelity for the Kronecker-delta
initial state defined in Eq.~(\protect\ref{Kron}) and the Gaussian-type
initial state defined in Eq.~(\protect\ref{Gauss}) with $\protect\alpha=5$,
respectively, for different values of the transverse field $T$ and
longitudinal field $F$. Unless otherwise specified, $F=0.1$ and $N_{\mathrm{A%
}}=20$. The insets in (a1) and (b1) show the corresponding results for $%
F=0.2 $. Here, $T_{\mathrm{B}}=\protect\pi/F$ denotes the analytically
predicted period of the Bloch oscillation. Panels (a2) and (b2) [panels (a3)
and (b3)] show the corresponding local magnetization, defined in Eq.~(%
\protect\ref{m_j}), for the initial states considered in (a1) and (b1),
respectively, under the driving Hamiltonian with $T=0.1$ and $F=0.1$ ($F=0.2$%
). The results in (a1) and (b1) show that the fidelity exhibits a clear
periodic behavior consistent with the theoretical prediction. The
domain-wall position oscillates periodically with period $T_{\mathrm{B}}$,
while the oscillation amplitude is determined by the competition between $T$
and $F$. For the localized initial state, different values of $T$
significantly affect the oscillation amplitude. In contrast, the
Gaussian-type initial state is less sensitive to $T$ because its initial
domain-wall position is spatially extended.}
\label{Fig3}
\end{figure*}

\section{Wannier-Stark ladder}

\label{Wannier-Stark ladder}

We consider a transverse-field Ising model defined on a chain of $N_{\text{%
\textrm{A}}}$ sites. The Hamiltonian is decomposed into two parts, 
\begin{equation}
H_{\text{\textrm{eff}}}=H_{\mathrm{V}}+H_{\mathrm{T}},  \label{H_eff}
\end{equation}%
where $H_{\mathrm{V}}$\ represents the diagonal part of the Hamiltonian,%
\begin{equation}
H_{\mathrm{V}}=\sum_{j=1}^{N_{\text{\textrm{A}}}-1}\tau _{j}^{z}\tau
_{j+1}^{z}+F\sum_{j=1}^{N_{\text{\textrm{A}}}}\tau _{j}^{z},  \label{Hv}
\end{equation}%
with $F=\left( h_{1}-h_{2}\right) /2$, while $H_{\mathrm{T}}$\ describes the
off-diagonal transition term,%
\begin{equation}
H_{\mathrm{T}}=T\sum_{j=1}^{N_{\text{\textrm{A}}}}\left( \tau _{j}^{+}+\tau
_{j}^{-}\right) .  \label{HT}
\end{equation}%
Here, $H_{\mathrm{T}}$ couples different eigenstates of $H_{\mathrm{V}}$ by
flipping the pseudospins and therefore induces transitions between different
energy configurations. In the $\left\{ \tau _{j}^{z}\right\} $ basis, $H_{%
\mathrm{V}}$ and $H_{\mathrm{T}}$ correspond to the diagonal and
off-diagonal components of the matrix representation of the Hamiltonian,
respectively.

Here, we assume that $F,T\ll 1.$\ Under this assumption, the eigenstates of $%
H_{\mathrm{V}}$\ can be divided into multiple subspaces according to their
energies, which are primarily determined by the number of domain walls.
Within the single-domain-wall subspace, the low-energy excitations of $H_{%
\mathrm{V}}$ above a ferromagnetic background can be described by
domain-wall states. We consider a reference ferromagnetic state,%
\begin{equation}
\left\vert FM\right\rangle =|\bullet \bullet \bullet \cdots \bullet \rangle ,
\end{equation}%
where $\left\vert \bullet \right\rangle _{l}=\left\vert \uparrow \downarrow
\right\rangle _{l}$, $\left\vert \circ \right\rangle =\left\vert \downarrow
\uparrow \right\rangle _{l}$) denotes the eigenstate of $\tau _{l}^{z}$\
with eigenvalue $1$\ ($-1$), respectively, and we introduce a single
domain-wall excitation,%
\begin{equation}
\left\vert j\right\rangle =|\circ \cdots \circ \bullet \cdots \bullet
\rangle ,  \label{one-domain}
\end{equation}%
where the domain wall is located between sites $j$ and $j+1$. These states
form an effective basis for the propagation of a domain wall along the chain.

The Ising interaction term in $H_{\mathrm{V}}$ gives the same contribution
for all single-domain-wall states, whereas the longitudinal field term
generates a position-dependent energy. Specifically, $\left\langle
j\right\vert H_{\mathrm{V}}\left\vert j\right\rangle =E_{0}-2Fj$, where $%
E_{0}$ is an irrelevant constant. Therefore, the domain-wall excitation
experiences an effective linear potential, $V(j)=-2Fj$. This linear
potential is analogous to the tilted potential responsible for the
conventional Wannier--Stark effect. As a consequence, the domain-wall
eigenstates form an equally spaced energy ladder,%
\begin{equation}
E_{j}=E_{0}-2Fj,
\end{equation}%
known as the Wannier--Stark ladder.

Unlike conventional Bloch oscillations, where the linear potential is
externally imposed by a constant force in real space, the effective force in
the present system emerges from the imbalance between the longitudinal
fields of the two coupled Ising layers. Therefore, the Wannier--Stark ladder
arises from an intrinsic many-body mechanism associated with the emergent
pseudospin degrees of freedom.

When the term $H_{\mathrm{T}}$\ is switched on, the domain-wall states are
coupled through pseudospin-flip processes. For the domain-wall basis states $%
\left\vert j\right\rangle $, the action of $H_{\mathrm{T}}$ can be written
as 
\begin{equation}
H_{\mathrm{T}}\left\vert j\right\rangle =T\left\vert j-1\right\rangle
+T\left\vert j+1\right\rangle .
\end{equation}%
Here, we neglect the states containing multiple domain walls because they
are separated by large energy gaps. These relations show that the
transverse-field term only couples domain-wall states whose positions differ
by two lattice sites, corresponding to transitions between Wannier--Stark
levels $E_{j}$ and $E_{j\pm 1}$. Consequently, the effective Hamiltonian
takes the form%
\begin{equation}
H_{\mathrm{eff}}=T\sum_{j=1}\left( \left\vert j+1\right\rangle \left\langle
j\right\vert +\mathrm{H.c.}\right) -2F\sum_{j=1}j\left\vert j\right\rangle
\left\langle j\right\vert .
\end{equation}%
Here, an overall constant energy shift has been omitted since it does not
affect the dynamics. This Hamiltonian corresponds to the standard
tight-binding description of a Wannier--Stark ladder.

It is well known that the spectrum of $H_{\mathrm{eff}}$ is equally spaced
levels, which supports periodic dynamics \cite{Hartmann2004,Bloch1929,
Zener1934, Zhang2024magnetic, Ma2026dynamic}. Such Wannier-Stark ladders
have been realized in semiconductor superlattices and ultracold atoms \cite%
{Waschke1993,BenDahan1996,Wilkinson1996,Anderson1998}.

The eigenstates $\left\vert \psi _{n}\right\rangle $\ satisfying the Schr%
\"{o}dinger equation 
\begin{equation}
H_{\mathrm{eff}}\left\vert \psi _{n}\right\rangle =E_{n}\left\vert \psi
_{n}\right\rangle ,
\end{equation}%
are explicitly expressed in terms of the Bessel function $J_{m}$\ as%
\begin{equation}
\left\vert \psi _{n}\right\rangle =\sum_{m}J_{m-n}(\frac{T}{F})|m\rangle ,
\end{equation}%
with $E_{n}=-2nF$, where $J_{m-n}(x)$ denotes the Bessel function of the
first kind. Importantly, the energy spacing of the Wannier--Stark ladder,%
\begin{equation}
E_{n+1}-E_{n}=-2F,
\end{equation}%
is independent of the hopping amplitude $T$, while $T$ only determines the
spatial extent of the localized eigenstates.

The above effective description is obtained within the single-domain-wall
subspace. In the full Hilbert space, the transverse-field term may also
couple these states to other excitation sectors. However, such processes are
suppressed when the Wannier--Stark energy scale is well separated from other
excitation energies. Fig. \ref{Fig2}\ schematically illustrates the energy
level structure and the transition processes between them, showing that the
effective Hamiltonian remains valid under the condition $F,T\ll 1$. The
validity of the effective Hamiltonian will be verified by numerical
simulations of the domain-wall dynamics in the following section.

\begin{figure*}[tbh]
\centering
\includegraphics[width=1.0\textwidth]{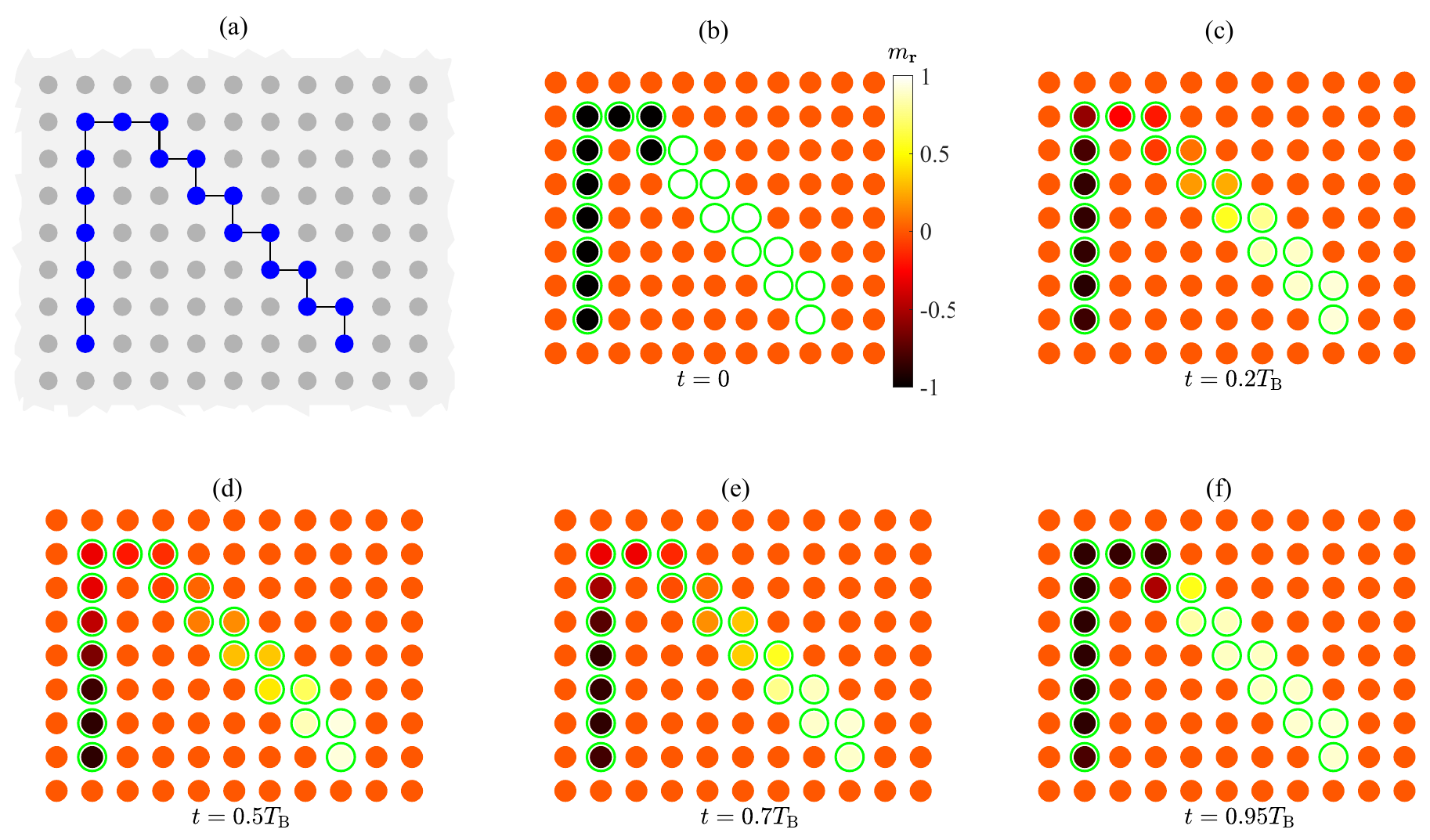}
\caption{(a) Schematic illustration of a one-dimensional chain described by
the effective Hamiltonian in Eq.~(\protect\ref{H_eff}), with $N_{\mathrm{A}%
}=20$ (blue filled circles), embedded in a square lattice. For the
Kronecker-delta initial state defined in Eq.~(\protect\ref{Kron}), panels
(b)--(f) show the local magnetization defined in Eq.~(\protect\ref{m_j}) at
several representative time points within one period, $T_{\mathrm{B}}=%
\protect\pi/F$. The other parameters are $F=0.1$ and $T=0.4$. These results
indicate that the magnetization approximately returns to its initial
distribution at $t=0.95T_{\mathrm{B}}$.}
\label{Fig4}
\end{figure*}

\begin{figure*}[tbh]
\centering
\includegraphics[width=1\textwidth]{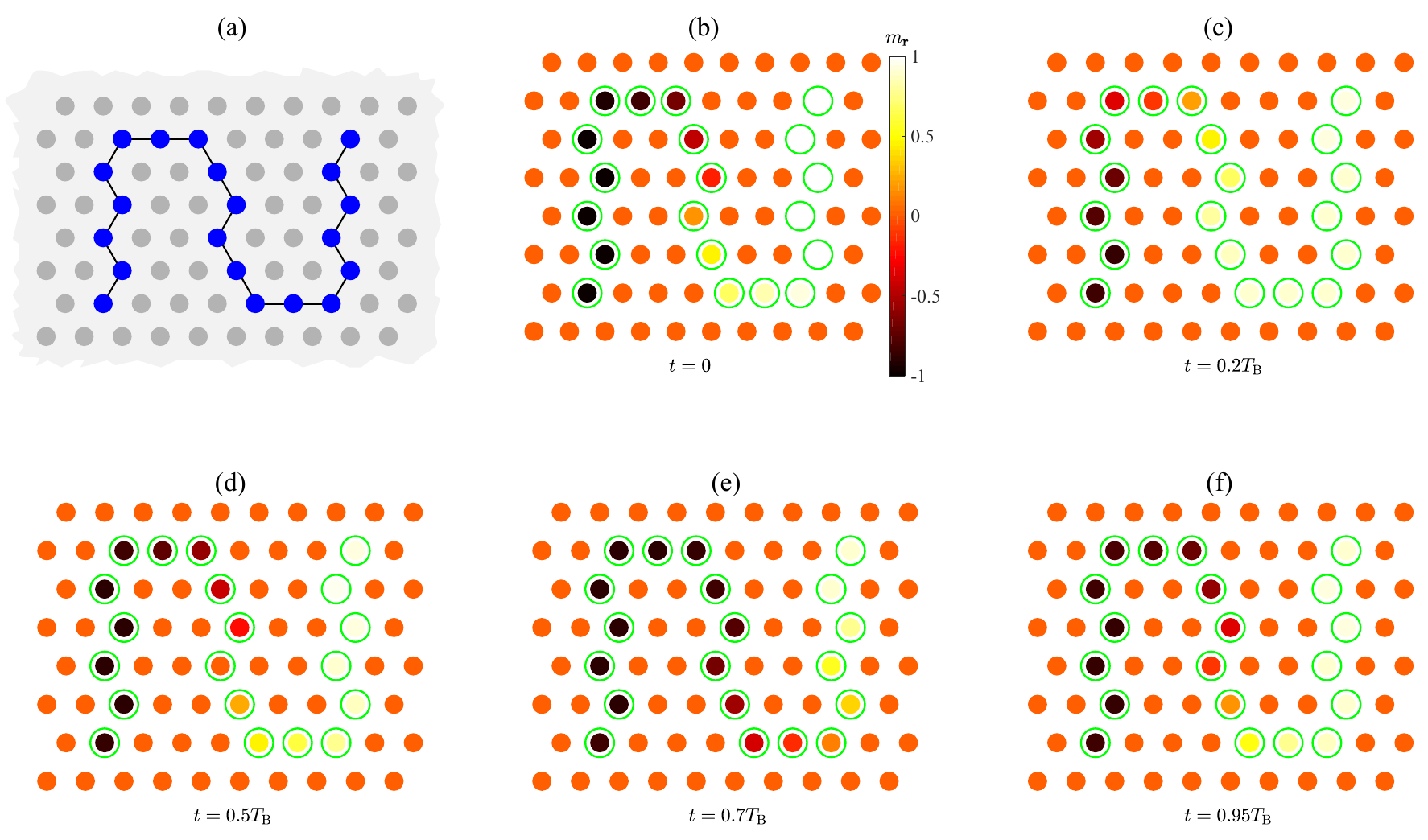}
\caption{(a) Schematic illustration of a one-dimensional chain described by
the effective Hamiltonian in Eq.~(\protect\ref{H_eff}), with $N_{\mathrm{A}%
}=20$ (blue filled circles), embedded in a hexagonal lattice. For the
Gaussian-type initial state defined in Eq.~(\protect\ref{Kron}) with $%
\protect\alpha=5$, panels (b)--(f) show the local magnetization defined in
Eq.~(\protect\ref{m_j}) at several representative time points within one
period, $T_{\mathrm{B}}=\protect\pi/F$. The other parameters are $F=0.1$ and 
$T=0.4$. These results indicate that the magnetization approximately returns
to its initial distribution at $t=0.95T_{\mathrm{B}}$.}
\label{Fig5}
\end{figure*}

\section{Bloch oscillations}

\label{Bloch oscillations}

The periodic dynamics of Wannier--Stark ladder systems, manifested as Bloch
oscillations, are well established. In each invariant sector, the time
evolution generated by the effective Hamiltonian in Eq. (\ref{H_eff}) is
described by the propagator 
\begin{equation}
K_{jj^{\prime }}\left( t\right) =\sum_{n}e^{-iE_{n}t}\left\langle j|\psi
_{n}\right\rangle \left\langle \psi _{n}|j^{\prime }\right\rangle .
\end{equation}%
Using the Wannier--Stark eigenstates $\left\vert \psi _{n}\right\rangle $\
and the equally spaced energy spectrum $E_{n}$,\ the propagator can be
expressed as 
\begin{equation}
K_{jj^{\prime }}\left( t\right) =\sum_{n}e^{i2ntF}J_{j-n}\left( \frac{T}{F}%
\right) J_{j^{\prime }-n}^{\ast }\left( \frac{T}{F}\right) .
\end{equation}%
Applying Graf's addition theorem for Bessel functions \cite%
{abramowitz1966handbook,holthaus1996localization}, this summation can be
evaluated analytically 
\begin{equation}
K_{jj^{\prime }}\left( t\right) =i^{\left( j-j^{\prime }\right) }e^{i\left(
j+j^{\prime }\right) Ft}J_{j-j^{\prime }}\left[ \frac{-2T}{F}\sin \left(
Ft\right) \right] .
\end{equation}%
This result explicitly demonstrates that the propagator is periodic in time, 
\begin{equation}
K_{jj\prime }(t+T_{B})=K_{jj\prime }(t),
\end{equation}%
with the Bloch period%
\begin{equation}
T_{\text{\textrm{B}}}=\frac{\pi }{F}.
\end{equation}%
Therefore, the emergent linear potential generated by the imbalance between
the longitudinal fields of the two Ising layers gives rise to coherent Bloch
oscillations of domain-wall excitations.

To investigate the dynamical behavior, we consider the time evolution of two
representative initial states in the single-domain-wall subspace, 
\begin{equation}
\left\vert \psi (0)\right\rangle =\sum\nolimits_{l}d_{l}\left\vert
l\right\rangle .
\end{equation}%
The first is a state localized at a specific domain-wall position 
\begin{equation}
d_{l}=\delta _{l,l_{0}},l_{0}=N_{\mathrm{A}}/2,  \label{Kron}
\end{equation}%
which corresponds to a delta-function initial state. The second is a
Gaussian wave packet, 
\begin{equation}
d_{l}=\Omega ^{-1}e^{il\pi /2}e^{-(l-N_{\text{\textrm{A}}}/2)^{2}/\alpha
^{2}},  \label{Gauss}
\end{equation}%
where $\Omega $\ is the normalization factor. For finite systems, we perform
exact diagonalization to obtain the time evolution of the wave function. To
characterize the Bloch oscillations, we calculate the fidelity, 
\begin{equation}
F\left( t\right) =\left\vert \left\langle \psi \left( 0\right) |\psi \left(
t\right) \right\rangle \right\vert ^{2},  \label{F}
\end{equation}%
and the spatial distribution of the domain-wall excitation. In the original
spin representation, the latter can be monitored through the local
magnetization,%
\begin{equation}
\left\langle m_{\mathbf{r}}(t)\right\rangle =\frac{1}{2}\left\langle \psi
\left( t\right) \right\vert \left( \sigma _{\mathbf{r},1}^{z}-\sigma _{%
\mathbf{r},2}^{z}\right) \left\vert \psi \left( t\right) \right\rangle .
\label{m_j}
\end{equation}

Numerical results for several representative parameter sets are shown in
Fig. \ref{Fig3}, where clear Bloch oscillations of the domain-wall
excitations are observed. For both the delta-function initial state and the
Gaussian wave-packet initial state, the oscillation period is found to be $%
\pi /F$, in excellent agreement with the analytical prediction obtained from
the Wannier--Stark ladder spacing, $\Delta E=2F$. In addition, we present
the time-dependent local magnetization of a selected one-dimensional active
chain embedded in lattices with different geometries in Figs. \ref{Fig4} and %
\ref{Fig5}. These numerical results confirm that the imbalance between the
longitudinal fields of the two Ising layers acts as an emergent force for
the domain-wall excitations, giving rise to coherent Bloch oscillations
without the need for a conventional tilted potential. They also validate the
effective Wannier--Stark description derived from the energy-level structure
shown in Fig. \ref{Fig2}.

\section{Summary}

\label{conclusions}

In summary, we have investigated domain-wall dynamics in two coupled Ising
layers with asymmetric longitudinal fields. By introducing rung pseudospin
operators, we revealed a reducible SU(2) structure of the local Hilbert
space. This emergent pseudospin structure naturally leads to Hilbert-space
fragmentation, where each fragmented sector is governed by an effective
transverse-field Ising Hamiltonian. The interlayer Heisenberg coupling acts
as an effective transverse field, while the imbalance between the
longitudinal fields of the two layers generates an additional longitudinal
potential for the emergent pseudospins. Within the effective one-dimensional
sectors, we demonstrated that domain-wall excitations experience a linear
potential induced by the longitudinal-field imbalance. This emergent
potential gives rise to a Wannier--Stark ladder with equally spaced energy
levels, providing a mechanism for coherent Bloch oscillations without
applying a conventional tilted potential. Finally, numerical simulations of
both localized and Gaussian wave-packet initial states confirmed the
analytical predictions and demonstrated robust Bloch oscillations of
domain-wall excitations. Our results provide a distinct mechanism for
generating Wannier--Stark physics through internal many-body degrees of
freedom rather than externally imposed forces.

\acknowledgments This work was supported by the National Natural Science
Foundation of China (under Grant No. 12374461).

\bibliography{BO_Ladd_References.bib}

\end{document}